\documentclass[10pt]{article}
\usepackage{epsfig}
\usepackage{graphicx}
\usepackage{amsmath}
\usepackage{amssymb}
\usepackage{subfig}
\begin{document}
\title{Statistical Inference of Functional Connectivity in Neuronal Networks using Frequent Episodes}

\author{
Casey Diekman\footnote{Department of Industrial \& Operations Engineering, Ann Arbor, MI 48109, University of Michigan, email: diekman@umich.edu},
Kohinoor Dasgupta\footnote{Department of Statistics, University of Michigan, Ann Arbor, MI 48109, email: kohinoor@umich.edu},
Vijay Nair\footnote{Department of Statistics, University of Michigan, Ann Arbor, MI 48109, email: vnn@umich.edu},\\
P.S. Sastry\footnote{Department of Electrical Engineering, Indian Institute of Science, Bangalore, India 560012, email: sastry@ee.iisc.ernet.in},
K.P. Unnikrishnan\footnote{General Motors R \& D Center, Warren, MI 48090, email: k.p.unnikrishnan@gmail.com}
}

\maketitle
\begin{abstract}
Identifying the spatio-temporal network structure of brain activity from multi-neuronal
data streams is one of the biggest challenges in neuroscience. Repeating patterns of
precisely timed activity across a group of neurons is potentially indicative of a
microcircuit in the underlying neural tissue. Frequent episode discovery, a temporal data
mining framework, has recently been shown to be a computationally efficient method of
counting the occurrences of such patterns. In this paper, we propose a framework to
determine when the counts are statistically significant by modeling the counting process.
Our model allows direct estimation of the strengths of functional connections between
neurons with improved resolution over previously published methods. It can also be used
to rank the patterns discovered in a network of neurons according to their strengths and
begin to reconstruct the graph structure of the network that produced the spike data. We
validate our methods on simulated data and present analysis of patterns discovered in
data from cultures of cortical neurons.\\
\\
\textbf{keywords:} event sequences, spike trains, multi-electrode array, microcircuits, temporal data mining, frequent episodes, non-overlapped occurrences, statistical inferences
\end{abstract}

\section{Introduction}
\label{sec:intro}

A brain tissue is composed of many neurons which function and
interact with each other by generating a time sequence of
characteristic electrical pulses known as action potentials or
spikes \cite{Brillinger:Max}. These time sequences of spikes generated by a
set of neurons is referred to as multi-neuronal spike train
data. Such data contain the stochastic firing (or spike
generating) events of individual neurons as well as spiking
activity due to coordinated functioning of a group of neurons
which are functionally interconnected. Analyzing the
multi-neuronal data to identify the spatio-temporal network
structure of the functional connectivity of the neurons
underlying a specific brain activity is one of the biggest
challenges in neuroscience.

Multi-electrode array (MEA) recordings have become a standard
neuroscience tool in recent years, leading to large collections
of multi-neuronal data, with each collection involving spikes from many
neurons (see Figure 7 for an example of MEA data). It is critical that we are able to analyze data from
these multi-channel studies in order to understand how groups
of neurons in a given region of the brain act together in
response to external stimuli or internal mental processes.
However, there is limited ability currently to process such
large amounts of spatial and temporal data quickly and
efficiently and analyze them to make the appropriate inference
\cite{Brown:spike}. Many of the common techniques,
such as cross-correlograms, cross-intensity functions, and
joint peri-stimulus time histograms, are aimed at analyzing
neurons pairwise or at most a small number of neurons at a time \cite{Brown:spike}.

Frequent episode discovery \cite{Mannila:freqepi}, a popular framework for temporal data mining, has
recently been proposed as a method for characterizing temporal
firing patterns from multi-neuronal spike train data \cite{Patnaik:inf}. Temporal data mining deals with
the analysis of time-stamped symbolic (categorical) data
streams, specifically a sequence of time stamped events of a
given type. In the analysis of multi-neuronal data, the events
are spikes, and the event type would be the neuron that
generated the spike. We are interested in \emph{serial
episodes} given by an ordered collection of event types, namely
certain events occurring in the prescribed order (such as a firing
of neuron $A$ followed by firing of neuron $B$). These episodes can
also be specified with temporal constraints, for example,
specific time delays between the firings of the two neurons.
Patnaik, Sastry and Unnikrishnan (2008) have used temporal data
mining methods to compute the frequent episodes of interest:
repeated occurrences of a precisely timed pattern of spiking
activity across a collection of neurons. A review of the
approach is given in Section 2.

A major challenge in implementing the algorithms in practice is
that it is computationally expensive to compute the complete
set of frequent episodes of all order. We need
statistically-defined thresholds that allow us to delineate the
unimportant ones (random or weak connectivity) from the
significant ones, so that only counts above pre-specified
thresholds need to be counted. Section 3 of the paper develops
the statistical distribution of the number of frequent episodes
as well as that of the number of `non-overlapping' episodes
(defined in Section 2). This is an important result that allows
practical implementation of the temporal data mining algorithm.
Earlier work \cite{Sastry:cond} used recurrence
relations to compute the mean and variance of these quantities.
The results here provide a complete characterization of the
distribution which allows us to compute all moments of the
distribution as well as obtain good approximations to the
quantiles. Section 4 discusses how the results can be used for
statistical inference about the connectivity strength of the
neurons (the strength of the edges of the network). Section 5
shows an application of the results to simulated multi-neuronal
data, while Section 6 demonstrates the usefulness of the
results on a real neuronal data set. The method presented here employs
fixed time delays between neurons in the firing patterns and
considers only serial patterns. We conclude the paper with a discussion of how the method can
be extended in Section 7.

\section{Frequent episode discovery in spike train data}
\label{sec:episodes}

The notion that information may be encoded in the precise
timing of neuronal spikes, rather than only in firing rates,
was suggested by Hebb when he introduced the concept of
neuronal assemblies \cite{Hebb:org}. When activated, these ``Hebb
cell-assemblies'' would produce a characteristic
spatio-temporal pattern of spikes. Discovering frequent
episodes is thus akin to finding pieces of a neuronal assembly
which are functionally connected. Subsequent analysis of the
set of frequent episodes discovered can lead to a
reconstruction of the assembly as a graph of its functional connections.

The most popular method for detecting repeated occurrences of
precise patterns of spikes is the two-tape algorithm of Abeles
and Gerstein (1988), which is based on correlations between
time-shifted spike trains of two neurons. As mentioned earlier,
most currently available statistical techniques can handle
computations involving only a small number of neurons
(e.g. 5). However, spatio-temporal groups or
``polychronous circuits'' involving as many as 25 neurons have been suggested based on modeling studies \cite{Izhikevich:Spike}\cite{Izhikevich:Poly}. The ability to
simultaneously record data from an increasing number neurons
necessitates the development of algorithms for inferring firing
patterns involving large number of neurons. The data-mining
method proposed in Patnaik et al (2008) is
an efficient technique for computing patterns involving many
neurons. In this section, we provide a brief overview of the
frequent episodes framework \cite{Mannila:freqepi}\cite{Laxman:disc} and the data-mining algorithm of Patnaik et al (2008).

The data to be analyzed are viewed as a time-ordered sequence
of events denoted by ${\cal D} = \{ (E_1, t_1), \cdots, (E_i,
t_i) \}$. Here, $E_i$ denotes the event type (which takes
values from a finite set) and $t_i$ denotes the time of
occurrence. The data is ordered so that $t_i \leq t_{i+1}$,
$\forall i$. For the spike train data, the event type of each
spike event would be the identity of the neuron that generated
that spike. Thus, by collecting together all spikes from all
neurons and ordering them in time, we get the data as an event
stream.

The temporal patterns of interest in this framework are called
{\em episodes}, which are, in general, partially ordered sets of
some of the event types. In this paper, we consider only the
so-called \emph{serial episodes}, or (short) ordered, sequences
of event types. For example, a serial episode $A \rightarrow B
\rightarrow C$ means the ordered event \emph{A followed by B followed by C}.
A serial episode is said to occur in a data stream if there is
a large enough number of events of the appropriate types in the
data that conform to the order imposed by the episode. For
example, consider the following data sequence:

\medskip
\noindent [(A,1), (D,3), (C,4), (A,5), (B,7), (B,8), (C, 11), (C,12), (F,12), (A,13), (B,15), (C,17)] \\
\medskip

\noindent This means means neuron $A$ fired at time unit 1, neuron $D$ fired at time unit 3, and so on.
If we consider the subset of events $<(A,5), (B,7), (C,11)>$,
there is an occurrence of $A \rightarrow B \rightarrow C$. There are three distinct
occurrences of this episode in the above data set. A serial episode
such as $A \rightarrow B \rightarrow C$, when
it occurs often in the spike data, would be an interesting
microcircuit. This is because it represents a specific time
ordered sequence of firing by the three neurons and such microcircuits may play a role in information processing in the brain.

Given a large set of multi-neuronal data, the computational
task is to automatically discover all {\em frequent} episodes.
It is not tractable to compute all distinct frequent episodes
in a large data set, so we will focus on non-overlapping
episodes. Two occurrences of an episode are said to be
non-overlapping if their corresponding time periods do not
overlap. For example, in the data sequence given above, the first
occurrence of $A \rightarrow B \rightarrow C$ overlaps with the
second one. (See Figure 1 for another illustration). A set of occurrences of an episode is said to be
non-overlapping if every pair is non-overlapping.

\begin{figure}[h]
\centering
\includegraphics[scale=0.5]{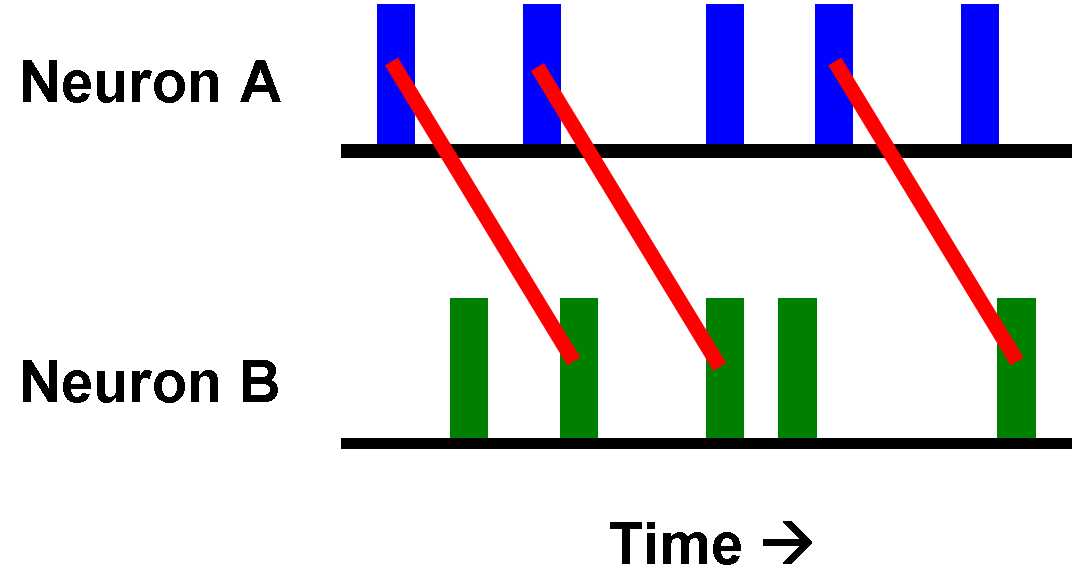}
\caption{Shown here are three occurrences of the episode $A[k]-B$, however at most 2 of the occurrences are nonoverlapping}
\end{figure}

Laxman et al (2005, 2007) developed efficient algorithms for
counting all non-overlapped occurrences of episodes. These algorithms use the Apriori-style
level-wise procedure that is common to many frequent pattern
discovery methods in data mining \cite{Agarwal:min}. At each level, say, $l$, we
use frequent episodes of size $(l-1)$ to generate candidate
$l$-node episodes and, through one pass over the data, count
the frequencies of all candidate episodes. Those counts over the threshold are considered frequent $l$-node episodes. We use finite-state automata based
counting procedures for obtaining the frequencies of all
candidate episodes. These methods are efficient when we are
counting only non-overlapped occurrences \cite{Laxman:fast}.

Apart from considerations of computational efficiency, the
count of non-overlapped occurrences as frequency is somewhat
natural in our application of spike train data analysis. If a
serial episode is to be interesting as (part of) a microcircuit
in the underlying neural tissue, then we would like it to
represent a chain of causative or triggerring events. A
frequent episode (based on count of nonoverlapping occurrences)
would be such a chain that occurs repeatedly at different
times.

However, for frequent serial episodes to be interesting in the
context of microcircuits in the neural tissue, we need
additional temporal structure in the episodes. Suppose neuron
$A$ strongly influences neuron $B$ which in turn influences
$C$. Hence we can expect $A \rightarrow B \rightarrow C$ to be
a frequent episode. However all such influences among neurons
are mediated by processes that have characteristic time delays on the order of milliseconds. To motivate why it is important to consider time delays when characterizing influence among neurons we will give a brief explanation of the physiology of neuronal communication. A neuron consists of a cell body, from which extends a long, thin structure called an axon. Axons form connections with other neurons at junctions called synapses. When a neuron fires a spike, the electrical impulse travels down the axon to the synapse, where chemicals (neurotransmitters) are released which diffuse across the synapse and may or may not cause the neuron on the other side of the synapse to fire a spike in response. The randomness of the diffusion process gives some insight into why conditional probabilities are a good way to measure the strength of a neuronal connection. Furthermore, the diffusion process takes some time, referred to as the synaptic delay. Axons can be rather lengthy, and so of!
 ten there is also a delay associated with propagation of the action potential down the axon.

Thus, if a spike from $B$ is viewed as being due to the
influence of $A$, then we should have the difference between
the times of the two spikes to be in some interval $[T,
T+\Delta T]$. So, for a frequent episode to represent
influences among underlying neurons, we should ensure that we
count only those occurrences of the episode which satisfy this
additional temporal constraint.

Hence, we consider serial episodes with inter-event time
constraints. We represent such an episode as $A[k_1]-
B[k_2]-C$. Here $k_1$ and $k_2$
represent the inter-event times that are required. For this
episode to occur, we need spikes from $A, B$ and $C$ in that
order and, in addition, the time between $A$ and $B$ should be
$k_1$ and the time between $B$ and $C$ should be $k_2$. In
multi-neuronal spike data, such serial episodes with
inter-event time constraints denote sequential firing by a
group of neurons with fixed delays between different neurons.
Such patterns, sometimes called precise firing sequences, are
important microcircuits, and automatically detecting such
sequences from spike data is very useful for understanding the
functional connectivity in the neural tissue.

In general, time delays between neurons are not strictly
constant. This can be taken care of by specifying the
inter-event time constraint as an interval rather than a single
number. However, in this paper, we will assume that the delays
are constant and when we specify a delay the units are milliseconds.

The frequent episode discovery algorithms mentioned earlier
have been generalized to the case of serial episodes with
inter-event time constraints \cite{Patnaik:inf}. This
algorithm can handle inter-event constraints specified as
intervals also. In addition, if the user specifies a set of
intervals (as all possible inter-event constraints), then the
algorithm can discover frequent episodes along with the most
appropriate inter-event time constraints. This is the algorithm
that we use in this paper in all our empirical results.

\section{Distributions of the Number of Occurrences of Serial Episodes}

For practical implementation of the algorithm discussed above, we need to set a threshold above which we deem episode counts to
be frequent. The rationale is that many of the counts are due to
random firings of the neurons, and we want to focus only on counts
that indicate true dependence. (We will restrict attention here to
positive dependence as the case of negative dependence or inhibition
is more difficult.) Additionally, using appropriate thresholds to determine frequent episodes of size $l-1$ allows us to gain computational efficiency by limiting the number of candidate patterns generated of size $l$. It is relatively straightforward to compute thresholds under the assumption that the neurons fire independently
of each other. It is more interesting (and useful) to compute
thresholds under appropriate notions of \emph{weak connectivity},
and look at counts larger than these thresholds, indicating moderate
to strong connectivity.  We will do this under the notion of
strength based on conditional probabilities introduced in Sastry and
Unnikrishnan (2008).

We will discretize time into bins of a fixed size $\Delta$. From now
on, when we say time $t$, it refers to the right end-point of the
bin, so that time $t$ refers to the bin $((t-1)\Delta, ~ t\Delta]$
for $t=1,2,...$ For simplicity, we will assume that the time delays
between the neurons are fixed, and denote this delay by $k$. Let
$P(B|A)$ denote the conditional probability that $B$ fires at time
$t+k$ given that $A$ fired at $t$. (We assume stationarity so that
this conditional probability is same for all $t$). Under
independence, $P(B|A) = P(B)$. Define $s = P(B|A)/P(B)$. Let $s_0$
be a user-defined threshold. We say that the influences of $A$ on $B$
is `weak' if $s < s_0$.

Now, we need to know the distribution of the number of frequent
episodes under the various degrees of strength so that we can
develop thresholds, or more formally statistical hypothesis tests
that can be used to detect the important episodes.

\subsection{Total Occurrences}

We start with the simpler case of the total number of occurrences of an
episode (including overlapped occurrences). Suppose we observe the
number of firings over a time period with $L$ bins, each of size
$\Delta$. Let $P_A$ be the probability that neuron $A$ fires in an
interval. Define $P_B$ similarly. Suppose we are interested in the
event $E = A[k]-B$, i.e., $B$ fires $k$ time units
after $A$ fires. Let $I_E(t) = 1$ if $E$ occurs at time $t$, i.e.,
$A$ fires at time $t$ and $B$ fires at time $t+k$. Let $P_E$ be the
probability of this event at any time $t$. Then, $P_E = P(B | A)
\times P(A)$, where $P(B | A) = P(A[k]-B|A)$. In other
words, $P_E = s P(B) \times P(A)$, where $s$ is the strength of the
connection.

Define $N$ as the total number of (possibly overlapping) events of
type $E$ during the observation period. Note that the possible
values of $N$ range from $0$ to $(L-k+1)$; any firing of $A$ that
occurs after time $t = L-k+1$ will not complete by time $L$.
Furthermore, under our model, the $I_E(t)$'s are independent and
identically distributed Bernoulli random variables with success
probability $P_E$. Hence $N$ has a binomial distribution with
parameters $(L-k+1)$ and $P_E$. In particular,

\begin{equation}
E(N)=(L-k+1)P_E
\end{equation}
and the variance is
\begin{equation}
Var(N)=(L-k+1)~P_E(1-P_E).
\end{equation}

\subsection{Non-Overlapping Occurrences}

However, we are more interested in the distribution of the number
of non-overlapping occurrences of type $E$, which we denote $M$. We obtain the distribution
of $M$ in this section.

Let $X_j, ~j=1,...,M$ be independent and identically distributed
binomial random variables with parameters $k$ as the number of
trials, and $P_E$ as the success probability. Further, the $X_j$'s
are independent of $M$.

\medskip

\noindent \emph{Proposition: $N$ has the same distribution as}
\begin{equation}
N ~=~ M + \sum_{j=1}^M X_j.
\end{equation}

\noindent \emph{Sketch of Proof:} Recall that $N$ is the total
number of events of type $E$ and that $M$ is the number of
non-overlapping occurrences. There are $k$ time units (bins) between
each of the $M$ non-overlapping occurrences. Consider the first such
interval, and let $X_1$ be the number of occurrences of $E$ where the
$A$ firings occur in this interval. Then, $X_1$ is a binomial random
variable with parameters $k$ (number of possible bins) and success
probability $P_E$. The independence of the $X_j$'s and with $M$
follows from the model assumptions.

\medskip

>From this, we can readily compute the moments of $M$. For example,
taking expectations of both sides of equation 3 and using the
independence assumption and known distributions of $N$ and $X_j$'s,
we get
\begin{equation}
E[M]~ = ~ \frac{E[N]}{1+E[X_1]}=\frac{L-k+1}{1/P_E+k}.
\label{eqn:exp}
\end{equation}
Similarly, the variance of $M$ is
\begin{eqnarray}
Var[M] & = & \frac{(L-k+1)P_E(1-P_E)}{(1+kP_E)^3}\\ & = &
\frac{Var[N]}{(1+kP_E)^3}
\end{eqnarray}

The variance of $M$ decreases as $k$ increases which is to be
expected. Figure 2 shows how the first two moments of $M$ vary as a
function of the parameters $L$, $k$, and $P_E$.

\begin{figure}[h]
\centering
\includegraphics[scale=0.5]{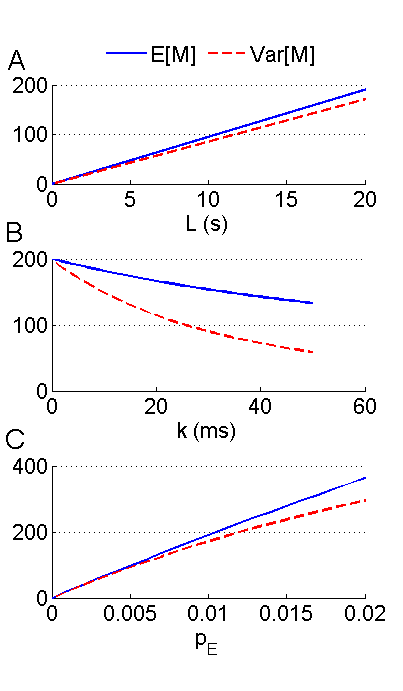}
\caption{Mean and variance of the non-overlapping count ($M$) of an episode as a function of \textbf{\emph{A:}} data length ($L$), \textbf{\emph{B:}} the span of the episode ($k$), and \textbf{\emph{C:}} the probability of the episode occurring ($P_E$).}
\end{figure}

While equation 3 gives us a complete characterization of the
distribution of $M$, it cannot be used to compute the actual
probability mass function or the cumulative distribution function.
However, we can obtain the moment generating function (MGF) of $M$ which can
be used to get all the moments. The MGF of $N$ is

\begin{eqnarray}
E[\exp(tN)] & = & E[\exp(t{(M+\sum_{j=1}^{M}X_j)})]\nonumber\\
 & = & E_M[E_X[\exp(t(M+\sum_{j=1}^{M}X_j))|M]]\nonumber\\
 & = &E[\exp(tM)(Q_E~+~P_E~\exp(t))^{Mk}]\nonumber\\
 & = &E[\exp(M(t+klog(Q_E+P_E \exp(t))))],\nonumber
 \label{eqn:mgf}
\end{eqnarray}
where $Q_E = 1-P_E$. Let $$s(t) = t + k \log(Q_E+P_E\exp(t)).$$ Note that this is a one-to-one and strictly
increasing function of $t$, so a unique inverse exists. Denote this inverse function by
$g(s)$. The function $g(s)$ has to be solved numerically. Then, the MGF of $M$ can be expressed as
$$E[\exp(sM)]=E[\exp(g(s)N)]=(Q_E+P_E\exp(g(s))^{L-k+1}.$$

We can get the $k$th moment of the
distribution of $M$ by differentiating the MGF $k$ times w.r.t. to $s$ and
evaluating it at $s=0$. The information about moments can be used to
numerically compute the probability mass function or cumulative
distribution function of $M$ to a good degree of approximation.

In this section, we will focus on the adequacy of a normal
approximation of $M$. Table 1 shows the skewness and kurtosis of $M$
calculated for a particular choice of $L$ and $k$ and a range of $P_E$ values. We have
subtracted 3 from the kurtosis for easy comparison with the normal
distribution. Note that these values are all quite close to zero
(except for the skewness for $P_E = 0.02$). We also compared the actual distributions of $M$ obtained using simulation with the
corresponding normal distribution (matching the mean and variance).
Specifically, we simulated 1,000 replications of $A$ and $B$ as
independent with 20Hz firing rates, and also with dependence of $B$ on $A$ ($P_{B|A}$=0.1). Figure 3 shows the normal quantile-quantile plots which should be
linear if the normal distribution is correct. We conclude from this
that, at least for the range of parameters considered here, the
normal approximation is quite reasonable. In future work we will study the adequacy of the normal
approximation and its refinement further. When the
normal approximation is reasonable, one can directly use it to
compute appropriate quantiles which can then be used as thresholds
for the non-overlapping counts.

\begin{table}[h]
\centering
\caption{Skewness and kurtosis of $M$}
\begin{tabular}{|c|c|c|c|c|c|l|} \hline
$P_E$ & skewness & kurtosis\\ \hline
0.02 & 0.1502 & 0.0206 \\ \hline
0.04 & 0.1007 & 0.0081 \\ \hline
0.08 & 0.0637 & 0.0021 \\ \hline
0.1 & 0.0537 & 0.0009 \\ \hline
0.2 & 0.0271 & -0.0012 \\ \hline
0.3 & 0.0140 & -0.0017 \\ \hline
0.4 & 0.0056 & -0.0019 \\ \hline
0.5 & -0.0004 & -0.0020 \\ \hline
0.6 & -0.0050 & -0.0021 \\ \hline
0.7 & -0.0086 & -0.0021 \\ \hline
0.8 & -0.0116 & -0.0022 \\ \hline
0.9 & -0.0141 & -0.0022 \\
\hline\end{tabular}
\end{table}

\begin{figure}[h]
\centering
\includegraphics[scale=0.4]{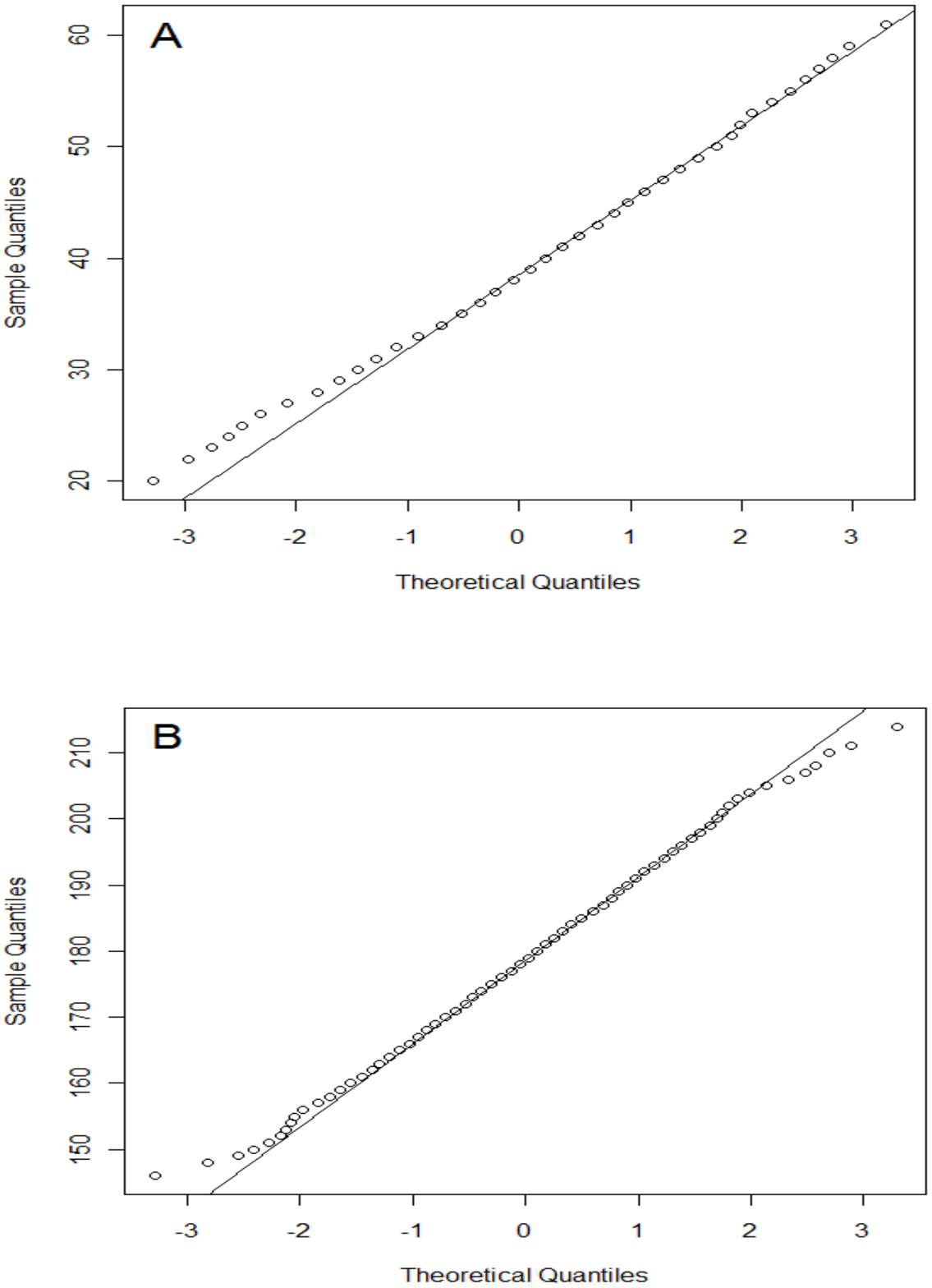}
\caption{Normal quantile-quantile plots of ($M$) for $L$=100s, $k$=50ms, and firing rate of $A$ as 20Hz. \textbf{\emph{A:}} $A$ and $B$ independent with firing rate of $B$ as 20Hz ($P_{B|A}$=0.02) \textbf{\emph{B:}} $B$ dependent on $A$ ($P_{B|A}$=0.1)}
\end{figure}

\label{sec:no}

\section{Inferring connection strengths}

A related quantity that is of even more interest than the count is the
connectivity strength $s$. In this section, we will consider
inference for the related quantity of interest $P_E$.

Suppose we conduct an experiment and observe the frequency
counts $M_j$ for various episodes of interest $E_j$, $j=1,2,...J$.
For any given event $E$, we can invert equation \ref{eqn:exp} to
estimate $P_E$ as
\begin{eqnarray}
\hat P_E=(\frac{L-k+1}{M}-k)^{-1}.
\end{eqnarray}
If the neurons $A$ and $B$ fire independently, $P_E$ will be simply
$P_E=P(A)\times P(B)$. However, we can estimate $P_E$ for any type of
dependence structure since we do not require independent firing of
$A$ and $B$ in our model. So in general, $P_E = P(B|A)\times P(A)$. Once we have an estimate of $P_E$, we can use it and an estimate of $P(A)$ to ultimately obtain an estimate of the conditional probability $P(B|A)$.

A confidence interval or hypothesis test for $M$ can
be directly inverted to get a confidence interval or test of
hypothesis for $P_E$. To demonstrate this, we simulated two neurons $A$ and $B$ with a variety of connections
strengths as shown in Table 2. For each simulation, we obtained a
count of the non-overlapping occurrences of the episode $A[k]-B$,
which we used to estimate $P_E$. The normal approximation for the
95\% confidence limits for $M$ was mapped to the corresponding 95\%
confidence limits for $P_E$ and hence the strength $s$. Table 2 shows the empirical
coverage probabilities which are very close to the true value of
$0.95$. Figure 4 shows the confidence intervals obtained for 100 replications at 2
different connections strengths.

\begin{table}[h]
\centering
\caption{ Estimated strength of functional connection $A[k]-B$ (1000 replications)}
\begin{tabular}{|c|c|c|l|} \hline
& &  Sample coverage \\
$P_{B|A}$ & $\hat{P}_{B|A}$ & probability of $\hat{P}_{B|A}$\\ \hline
0.02 & 0.0197 & 0.942  \\ \hline
0.04 & 0.0394 & 0.942 \\ \hline
0.08 & 0.0784 & 0.933 \\ \hline
0.1  & 0.0982 & 0.938 \\ \hline
0.2  & 0.1964 & 0.934 \\ \hline
0.3  & 0.2952 & 0.934 \\ \hline
0.4  & 0.3935 & 0.922 \\ \hline
0.5  & 0.4924 & 0.948 \\ \hline
0.6  & 0.5911 & 0.929 \\ \hline
0.7  & 0.6912 & 0.924 \\ \hline
0.8  & 0.7912 & 0.951 \\ \hline
0.9  & 0.8911 & 0.947 \\
\hline\end{tabular}
\end{table}

\begin{figure}[h]
\centering
\includegraphics[scale=0.4]{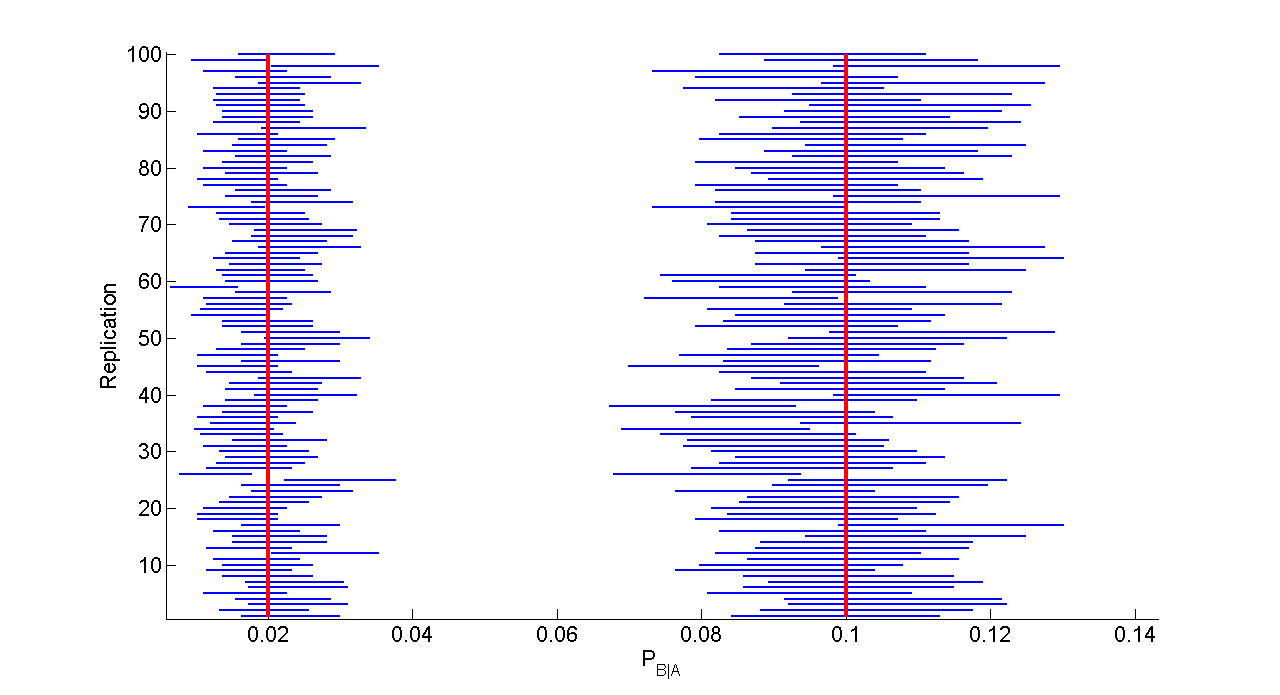}
\caption{95\% confidence intervals for estimate of $P_{B|A}$ based on a single count of $M$ (100 replications) for ${P}_{B|A}$=0.02 ($A$ and $B$ independent) and ${P}_{B|A}$=0.1}
\end{figure}

The event of interest can be an arbitrary $n$-node pattern of neuron firings. The derivations of expectation and variance of $M$ can also be extended to $n$-node patterns such as a 3-node episode
$A[k_1]-B[k_2]-C$. The expressions of expectation and variance of $M$ remain
unaltered, and the expression of $P_E$ depends on whether the neurons
fire independently. For a 3-node episode, the expression of $P_E$
becomes: \\

\noindent $P_E = P_A*P_B*P_C$, if $A$,$B$,$C$ are independent\\
$P_E = P_A*P_{B|A}*P_{C|B,A}$, if $B$ and $C$ are dependent on $A$  \\
$P_E = P_A*P_{B|A}*P_{C|B}$,  if $A$ and $C$ fire independently.\\

\section{Application to a simulated neuronal network}

We now demonstrate that our counting algorithm, combined with its statistical characterization, can be used to efficiently discover patterns and estimate their strength in a more complicated network of neurons. We simulate 25 neurons, each having an independent firing rate of 20Hz, with 5 different 3-neuron patterns embedded as shown in Figure 5. (Details of the neuronal simulator used in this paper are described in \cite{Patnaik:inf}\cite{Sastry:cond}.) As discussed in Section 2, we employ a level-wise mining scheme. At each level, we obtain an estimate of $\hat{P}_E$ for every episode counted, which we then use to determine if that episode is frequent or not.

\begin{figure}[h]
\centering
\includegraphics[scale=0.4]{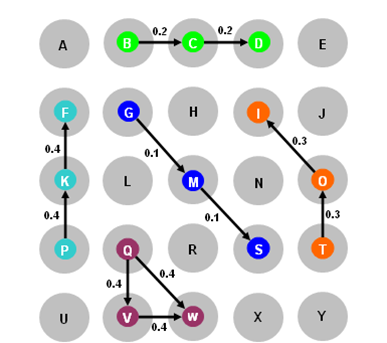}
\caption{Five 3-neuron patterns of various strength are embedded in a network of 25 neurons. All connection delays are 5 ms, except for the connection $Q-W$ which has a delay of 10 ms. The conditional probability for each connection is displayed. We simulate one dataset from this network, and then use our counting algorithms and statistical model to recover the network connectivity structure based on the spike data alone.}
\end{figure}

\subsection{Connection strength ratio}
We start at the 2-node level and count the nonoverlapped occurrences of all possible 2-node episodes. We mine using three different inter-event time constraints (5, 10, and 15 ms). Using equation 9, we map these counts to a $\hat{P}_E$ for each episode. We also calculate what the probability of each episode occurring would be if all the neurons participating in the episode were firing independently. For a 2-node episode such as $A[k]-B$, this probability would simply be $\hat{P}_A\hat{P}_B$. We estimate $\hat{P}_A$, $\hat{P}_B$ from the data as their respective 1-node counts divided by the total data length $L$. We then calculate the \textbf{connection strength ratio}, $s$, for this episode as $\frac{\hat{P}_E}{\hat{P}_A\hat{P}_B}$. If $s$ is significantly greater than one, it indicates there is an excitatory functional connection between neurons $A$ and $B$ with delay $k$. We will use this strength ratio as our threshold for mining. To illustrate, say we are only interested!
  in connections that are at least twice as strong as independence. Thus only episodes with a strength ratio of 2 or greater are deemed frequent, and only these frequent 2-node episodes are used to generate 3-node candidates for counting. This is much more efficient than counting all possible 3-node permutations (which would be 13,800 in this example). Table 3 shows all the 2-node episodes that would be considered frequent (strength ratio > 2) for one particular simulation. The 11 true 2-node connections embedded (see Figure 5) are all deemed frequent. The 2-node patterns ranked 10th, 13th, and 14th have ratios greater than 2 but are not in fact true connections. We will now show how we can use the counts of 3-node patterns to eliminate these false 2-node connections.

\begin{table}[h]
\centering
\caption{Discovered 2-Neuron Patterns}
\begin{tabular}{|c|c|c|c|l|} \hline
Rank & Pattern & Count & Strength Ratio \\ \hline
1 & Q[10]-W & 1158 & 17.2 \\ \hline
2 & V[5]-W & 1473 & 15.2 \\ \hline
3 & Q[5]-V & 778 & 15.2 \\ \hline
4 & P[5]-K & 713 & 14.5 \\ \hline
5 & K[5]-F & 978 & 13.8 \\ \hline
6 & T[5]-O & 537 & 11.3 \\ \hline
7 & O[5]-I & 712 & 11.2 \\ \hline
8 & B[5]-C & 378 & 8.6 \\ \hline
9 & C[5]-D & 435 & 8.6 \\ \hline
10 & P[10]-F & 301 & 5.5 \\ \hline
11 & G[5]-M & 179 & 4.6 \\ \hline
12 & M[5]-S & 188 & 4.3 \\ \hline
13 & T[10]-I & 166 & 3.3 \\ \hline
14 & B[10]-D & 106 & 2.4 \\
\hline\end{tabular}
\end{table}

\subsection{False edge elimination}
We obtain the 3-node pattern counts as shown in Table 4. We will take the 3-node pattern $P[5]-K[5]-F$ as an example, for which the count is 286. At the 2-node level, we had found 301 occurrences of the pattern $P[10]-F$ and it was deemed frequent. However we now see that this episode may have appeared to be frequent because $P$ and $F$ are the first and last nodes of the 3-neuron linear chain $P[5]-K[5]-F$ and perhaps there is not a true $P[10]-F$ connection. We can test for this by subtracting the 3-node pattern count of $P[5]-K[5]-F$ from the 2-node pattern count $P[10]-F$, leaving us with 15 occurrences. We then map this modified count to a new $\hat{P}_E$, and take the ratio of this $\hat{P}_E$ over what we would expect the probability of having a $P[10]-F$ occurrence without a $K$ in the middle to be if $P$, $K$, and $F$ were independent. This probability is calculated as $\hat{P}_P\hat{P}_F(1-\hat{P}_K)$, where the $\hat{P}_i$ are again estimated from the respective 1!
 -node counts in the data. The strength ratio comes out to less than 1, and we conclude that $P[10]-F$ is a false edge. On the other hand, applying this same procedure to test the connection $Q[10]-W$ (by subtracting the counts of $Q[5]-V[5]-W$) results in a strength ratio greater than 2, leading us to correctly conclude that $Q[10]-W$ is indeed a true connection.

This demonstrates how we can use the higher order frequent episode counts to prune lower order connections. This enables us to discover the network structure or graph that most likely generated the spike data. We will expand upon this form of graph discovery in our future work.

\begin{table}[h]
\centering
\caption{Discovered 3-Neuron Patterns}
\begin{tabular}{|c|c|l|} \hline
Pattern & Count \\ \hline
Q[5]-V[5]-W & 732 \\ \hline
P[5]-K[5]-F & 286  \\ \hline
T[5]-O[5]-I & 145 \\ \hline
B[5]-C[5]-D & 75 \\ \hline
G[5]-M[5]-S & 10 \\
\hline\end{tabular}
\end{table}

\section{Application to cultured cortical neurons}
Multi-electrode arrays (MEAs) are a common technique for recording the spiking activity of many neurons simulataneously. Cultures of dissociated neurons can be grown on an MEA dish for an extended period of time. When the cells are first placed on the dish, there are presumably no connections between the neurons as all the connections are destroyed during the dissociation process. However, over time the neurons potentially form new synapses, or connections, with other neurons in the culture resulting in microcircuits. These microcircuits may produce stereotypical patterns of spikes. Here we demonstrate that our methodology is capable of detecting such patterns, and therefore microcircuits, in data from a neuronal culture. Furthermore since we can estimate the strength of the connections we can track the development of the circuit as it evolves. This is of great interest in the neuroscience community as \emph{in vivo} microcircuits may be involved in cognitive functions such !
 as learning and memory formation.

We demonstrate the usefulness of our methodology by analyzing the recordings from cultures of rat cortical neurons described in \cite{Wag:bmc}. In that study an array of 59 electrodes was used to record from neurons in culture over a period of weeks. This dataset was also analyzed in Rolston et al (2007), where precisely timed spiking patterns were found using the two-tape algorithm of \cite{Abeles:det}. For example, in Figure 2 on page 297 of \cite{Rolston:Prec}, those investigators report a repeating pattern of across the 4 electrodes labeled $86-87-77-84$ (these numbers correspond to the row and column number of the electrodes on the MEA). When mining the data using our methodology (allowing for synaptic delays between 0 and 5 ms), we discovered this same 4-node pattern (see Figure 7) frequently occuring in culture $2-1$. It represents a particulary robust circuit, and here we show that we can track the development of connections in this circuit as they evolve over time. !
 We first preprocess the data to remove bursts of spiking activity so that the firing rates in the culture are relatively constant throughout the approximately 30 minutes of recordings for each day. For the 2-node episode $87[1]-77$, we counted the occurrences and estimated the connection strength ratio from days \emph{in vitro} (DIV) 17 through 26 as shown in Figure 6B. We can see that the strength ratio for this episode increases dramatically after DIV 22, indicating the development of a functional connection between $87-77$. In Figure 6A we plot the spiking activity of $87$ and $77$. The culture media that provides nutrients to the neurons was replaced on DIV 21, causing a sharp increase in spiking activity on that day. Despite such large variations in firing rate over the course of days, our strength ratio is useful for day-to-day comparisons since it takes the overall firing rate of the neurons participating in the episode into account. In Figure 6B we also show several!
  connections, including $77-87$, that do not show an increase in stren
gth over this time period. These preliminary results suggest our methodology is a promising way to track the development of a neuronal microcircuit.

\begin{figure}[h]
\centering
\includegraphics[scale=0.3]{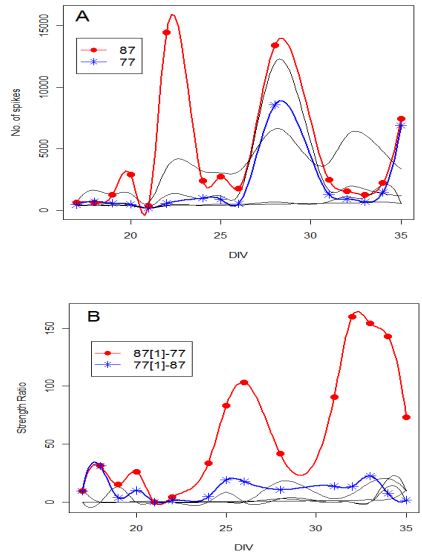}
\caption{Development of a microcircuit. \textbf{\emph{A:}} The spiking activity of neurons $87$ and $77$, as well as a few others (shown in black). \textbf{\emph{B:}} The strength ratio for $87[1]-77$ increases dramatically immediately following the culture media being replaced on DIV 21, indicating a functional connection. $77[1]-87$ and a few other connections (shown in black) do not show similar levels of increase in strength during this period of time.}
\end{figure}

\begin{figure*}[h]
\centering
\includegraphics[scale=0.5]{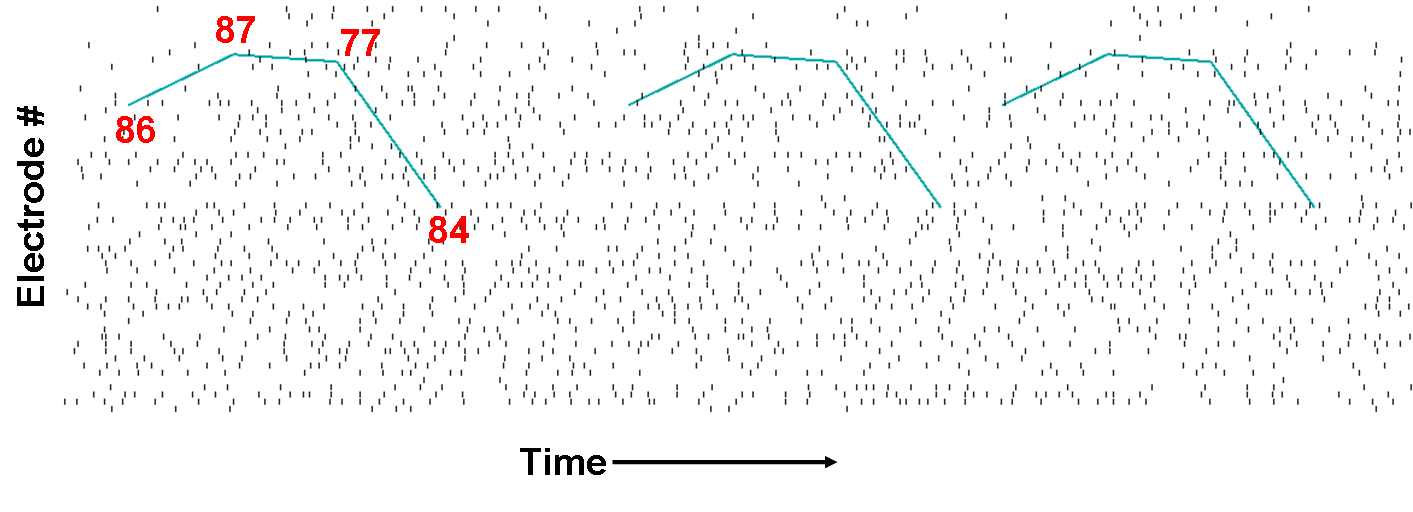}
\caption{Raster plot of spikes from cultured rat cortical neurons described in Wagenaar et al (2006) and Rolston et al (2007). There are 59 electrodes, and each tick mark represents a spike recorded on that particular electrode. Approximately 60 milliseconds worth of spikes from culture $2-1$ after 35 days \emph{in vitro} are displayed. Within these 60 milliseconds, there are 3 non-overlapping occurrences of the episode $86[5]-87[5]-77[5]-84$.}
\end{figure*}

\section{Discussion}

In this paper we have presented a datamining approach for the problem of discovering functional connectivity patterns from multi-neuronal spike train data. Inferring the graphical structure of connections in a neural  tissue is an important and challenging problem in neuroinformatics requiring computationally efficient techniques.  We have used the datamining framework of frequent episodes with inter-event  time constraints \cite{Patnaik:inf} for discovering serial connectivity patterns.  The main contribution of this paper is a statistical analysis method that allows one to infer which episodes represent strong interactions among neurons. We characterize the strength of connection between a pair of neurons as the conditional probability that one neuron fires after a specified delay given that the other neuron has fired. This delay is usually caused by the axonal delay of spike propagation and the delay at synapses due to chemical diffusion. By building a statistical model f!
 or the number of non-overlapped occurrences of a serial episode, we showed that the frequency (number of non-overlapped occurrences) of an episode can be directly used to estimate the strength of connections among the neurons in a network, constituting the episode. We demonstrated the effectiveness of our method both on simulated neuronal spike data as well as on data collected from {\em in vitro} neural cultures of cortical neurons.

In most frequent pattern discovery methods of datamining, the frequency threshold is a critical parameter. Hence an important problem in frequent pattern discovery is to relate the frequency to some generic characteristics of the data source so that one can fix the frequency threshold automatically using sound statistical considerations. Our approach of using the conditional probability to characterize the strength of interaction between neurons is generic and would be useful in many temporal datamining contexts. In analyzing any symbolic time series data to unearth interesting patterns in the form of episodes, the motivation is that such patterns represent dependencies among the symbolic event types that are inherent to the data source. Thus, the conditional probability that one event type would appear in the data stream after some time given that another has appeared, would be a useful characterization for the strength of the dependencies that we want to unearth. Hence, we!
  feel that the statistical method presented in this paper will be relevant in many other temporal datamining applications.

In the statistical analysis presented here, we have assumed that the inter-event time constraints which represent the time delays in neuronal connections are constants. In general there would be variation in these delays. Also, since we do not have complete control over which specific neurons are recorded from, we may be seeing some connections mediated by more than one synapse. Hence, one very useful extension for the method proposed here is to take care of variable delays as inter-event time constraints. Currently, our statistical framework handles only serial episodes which represent simple chains of connected neurons. Another interesting problem would be to extend the statistical theory for the case of more general episodes which can represent the connectivity graphs. We will be addressing such extensions in our future work.

\section{Acknowledgments}
We thank Dr. Steve Potter for providing us with the culture data and Debprakash Patnaik for coding the neuronal simulator and counting algorithms used in this paper. The work reported here is partially supported by a project funded by General Motors R\&D Center, Warren at the University of Michigan, Ann Arbor and the Indian Institute of Science, Bangalore.


\begin{thebibliography}{10}

\bibitem{Abeles:det}
M.~Abeles and G.~L. Gerstein.
\newblock Detecting spatiotemporal firing patterns among simultaneously
  recorded single neurons.
\newblock {\em Journal of Neurophysiology}, 60(3):909--924, September 1988.

\bibitem{Agarwal:min}
R.~Agrawal and R.~Srikant.
\newblock Mining sequential patterns.
\newblock In {\em Eleventh International Conference on Data Engineering}, pages
  3--14. IEEE Computer Society Press, September 1995.

\bibitem{Brillinger:Max}
D.~R. Brillinger.
\newblock Maximum likelihood analysis of spike trains of interacting nerve
  cells.
\newblock {\em Biological Cybernetics}, 59:189--200, 1988.

\bibitem{Brown:spike}
E.~M. Brown, R.~E. Kass, and P.~M. Mitra.
\newblock Multiple neural spike train data analysis: state-of-the-art and
  future challenges.
\newblock {\em Nature Neuroscience}, 7(5):456--461, May 2004.

\bibitem{Hebb:org}
D.~O. Hebb.
\newblock {\em Organization of Behavior: A Neuropsychological Theory}.
\newblock John Wiley and Sons, New York, 1949.

\bibitem{Izhikevich:Poly}
E.~M. Izhikevich.
\newblock Polychronization: Computation with spikes.
\newblock {\em Neural Computation}, 18(2):245--282, February 2006.

\bibitem{Izhikevich:Spike}
E.~M. Izhikevich, J.~A. Gally, and G.~M. Edelman.
\newblock Spike-timing dynamics of neuronal groups.
\newblock {\em Cerebral Cortex}, 14(8):933--944, 2004.

\bibitem{Laxman:disc}
S.~Laxman, P.~S. Sastry, and K.~P. Unnikrishnan.
\newblock Discovering frequent episodes and learning hidden markov models: A
  formal connection.
\newblock In {\em IEEE Transactions on Knowledge and Engineering}, volume~17,
  pages 1505--1517, March 2005.

\bibitem{Laxman:fast}
S.~Laxman, P.~S. Sastry, and K.~P. Unnikrishnan.
\newblock A fast algorithm for finding frequent episodes in event streams.
\newblock In {\em Data Mining and Knowledge Discovery}, August 2007.

\bibitem{Mannila:freqepi}
H.~Mannila, H.~Toivonen, and A.~Verkamo.
\newblock Discovery of frequent episodes in event sequences.
\newblock In {\em Data Mining and Knowledge Discovery}, volume~1, pages
  259--289, September 1997.

\bibitem{Patnaik:inf}
D.~Patnaik, P.~S. Sastry, and K.~P. Unnikrishnan.
\newblock Inferring neuronal network connectivity from spike data: A temporal
  datamining approach.
\newblock {\em Scientific Programming}, 16(1):49--77, 2008.

\bibitem{Rolston:Prec}
J.~Rolston, D.~A. Wagenaar, and S.~M. Potter.
\newblock Precisely timed spatiotemporal patterns of neural activity in
  dissociated cell cultures.
\newblock {\em Neuroscience}, 148(1):294--303, 2007.

\bibitem{Sastry:cond}
P.~S. Sastry and K.~P. Unnikrishnan.
\newblock Conditional probability based significance tests for sequential
  patterns in multi-neuronal spike trains.
\newblock {\em arXiv}, 0808.3511[q.bio], August 2008.

\bibitem{Wag:bmc}
D.~A. Wagenaar, J.~Pine, and S.~M. Potter.
\newblock An extremely rich repertoire of bursting patterns during the
  development of cortical cultures.
\newblock {\em BMC Neurosci}, 7(11), 2006.

\end{thebibliography}
\end{document}